\documentclass{PoS}

\PoS{PoS(WC2004)038}

\title{A Remark on the Aharonov-Bohm Potential and a Discussion on the Electric
Charge Quantization}

\ShortTitle{The Aharonov-Bohm Potential and the Electric Charge Quantization}

\usepackage{graphicx}

\author{\speaker{F. A. Barone}\\
        CBPF, Rio de Janeiro, Brazil\\
        E-mail: \email{fbarone@cbpf.br}}

\author{J. A. Helay\"el-Neto\\
        CBPF, Rio de Janeiro, Brazil\\
        E-mail: \email{helayel@cbpf.br}}

\abstract{The purpose of this work is to stress on a mathematical requirement of
the Stokes' theorem that, naturally, yields a reassessment of the electric
charge quantization condition, which is, here, explicitly carried out in the
context of the Aharonov-Bohm set-up. We argue that, by virtue of an ambiguity
in the definition of the circulation of the vector potential, a modified
quantization condition comes out for the electric charge that opens the way for
understanding fundamental fractional charges. One does not, any longer, need to
rely on the existence of a magnetic monopole to justify electric charge
quantization.}

\FullConference{Fourth International Winter Conference on Mathematical Methods
in Physics\\
                 09 - 13 August 2004\\
                 Centro Brasileiro de Pesquisas Fisicas (CBPF/MCT), Rio de
Janeiro, Brazil}

\begin{document}

\section{Introduction}

After Dirac's seminal papers on magnetic monopoles \cite{Dirac},
there appeared several others in the literature
\cite{Olive,Mignaco}, all of them presenting quantization
conditions for the electric charge, but involving magnetic
monopoles as well. We aim here at reassessing charge quantization
in the absence of monopoles. To do that, we reconsider the
Aharonov-Bohm experiment on diffraction of charged particles.

In their celebrated papers of 1959 and 1961, Aharonov and Bohm
\cite{AB,revAB} considered an infinitely long solenoid inside
which we could have a uniform and constant magnetic field; outside
no magnetic field is present. Outside the solenoid, Aharonov and
Bohm took a vector potential in such a way that its circulation
around a closed path would give the magnetic flux through a
surface whose the boundary was the considered path.

In this work, we show that it is not conflicting with Stokes'
Theorem to take other choices to the vector potential outside the
solenoid, even if their circulations around a closed path do not
give the flux of the magnetic field through a surface whose
boundary is the considered path. If we adopt this possibility
(that is valid for the Classical Mechanics), in the context of
Quantum Mechanics, and slightly and conveniently modify the
potential chosen by Aharonov and Bohm, we are straight taken to a
quantization condition for the possible values of the electric
charges in the nature, with no need to appeal for the existence of
magnetic monopoles.

The modification we consider in the Aharonov-Bohm choice for the
vector potential can be interpreted as if we had taken an
Aharonov-Bohm-like potential as being a pure gauge field. The
quantization condition obtained is consistent with the quark
fractional charges and with the existence of anti-particles.

\section{Quantization condition}

The basic set-up of the Aharonov-Bohm effect is an infinitely long
solenoid of radius $R$ where, in its interior, we have a uniform
and constant magnetic field; in its exterior, a null magnetic
field. One considers that no particles, or wave functions, can
penetrate in the interior of the solenoid.

In cylindrical coordinates, the magnetic field in the whole space
can be written as
\begin{equation}
\label{defB}
{\bf B}(\rho,\phi,z)=\cases{{\bf B}_{I}=B{\hat z},&$\rho<R$\cr {\bf B}_{E}={\bf 0},
&$\rho>R$}\ ,
\end{equation}
where $\phi$ is the azimuthal angle, and we considered the
solenoid axis lying on the $\bf z$ axis. On the solenoid,
$\rho=R$, the field is not defined.

In the interior of the solenoid, we can have a vector potential
given by ${\bf A}_{I}=B\rho/2\ {\hat\phi}$. In the exterior of the
solenoid, we can choose any vector potential, ${\bf A}_{E}$, whose
rotational is zero. In the context of Classical Electrodynamics,
an interesting choice is a vector potential of the form
\begin{equation}
\label{defA}
{\bf A}(\rho,z,\phi)=\cases{{\bf A}_{I}(\rho,z,\phi)={B\rho\over 2}{\hat\phi}, \
\rho<R\cr\cr {\bf A}_{E}(\rho,z,\phi)={\gamma\over\rho}{\hat\phi}, \ \rho>R}\ .
\end{equation}
where $\gamma$ is a constant. In the exterior region, the potential (\ref{defA})
has the property that its integral along any circular path perpendicular to the $z$-axis,
and centered at the origin, does not depend on the path radius:
\begin{equation}
\label{propr}
\oint_{\Gamma}{\bf A}_{E}\cdot d{\bf\ell}=2\pi\gamma\ ,
\end{equation}
for any circular path $\Gamma$ in the exterior region (for $r>R$), where $d{\bf\ell}$
stands for the tangent vector to the curve.

\begin{figure}
\begin{center}
\includegraphics[width=12.6cm]{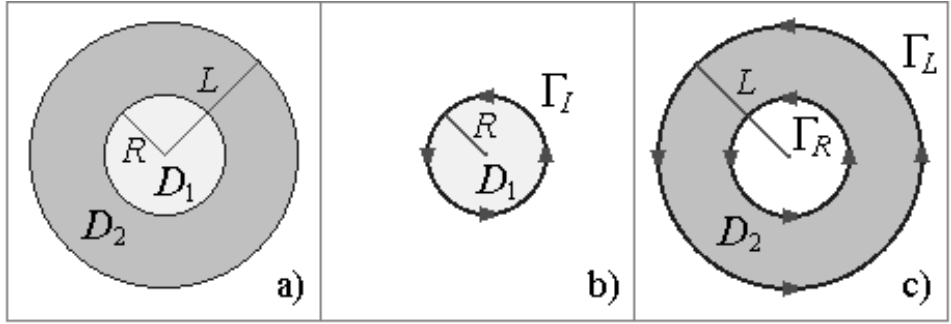}
\caption{a) Region $D$ divided into two others, $D_{1}$ and $D_{2}$, where the
vector potential is of class $C^{1}$. b) Region $D_{1}$ with the path $\Gamma_{I}$
in its boundary. c) region $D_{2}$ with the paths $\Gamma_{L}$ and $\Gamma_{R}$ in
its external and internal boundaries, respectively.}
\end{center}
\label{figura2}
\end{figure}

In order to calculate the flux of the magnetic field (\ref{defB})
through a disc $D$ of radius $L>R$ lying on the $xy$ plane and
centered at the origin, we can use Stokes' Theorem, but taking
into account two points: ({\it i}) the vector potential
(\ref{defA}) is not of class $C^{1}$ \cite{Marsden}, due to its
behavior at $\rho=R$, and ({\it ii}) we are using cylindrical
coordinates, which is not a one-to-one map on the region we want
to calculate the flux. To avoid the problem presented in ({\it
i}), we divide the disc $D$ into two regions: the first one,
$D_{1}$, is the internal region of the solenoid with $\rho<R$, and
the second one, $D_{2}$, is the external region of the solenoid,
with $R\leq\rho<L$, as indicated in figure (\ref{figura2}). Now,
the vector potential (\ref{defA}) is of class $C^{1}$ in both
regions, $D_{1}$ and $D_{2}$, and we can apply Stokes' Theorem in
each of them. The problem presented in ({\it ii}) can be easily
circumvented by using a Cartesian coordinate system in the
internal region of the solenoid, and dividing the external region
into two sectors, one of them with $0\leq\phi<\pi$ and the other
with $\pi\leq\phi<2\pi$. With this procedure, it can be shown that
point ({\it ii}) is irrelevant to this problem, and we can apply
Stokes' Theorem without taking that into account.

For $D_{1}$, we have the magnetic flux
\begin{equation}
\label{Phi1}
\Phi_{1}=\int\int_{D_{1}}{\bf B}_{I}\cdot d{\bf S}_{1}=\oint_{\Gamma_{I}}{\bf A}_{I}\cdot d{\bf\ell}=\pi BR^{2}\ ,
\end{equation}
where $\Gamma_{I}$ is the path around the boundary of the $D_{1}$
region, with positive orientation, as indicated in figure
(\ref{figura2}). The region $D_{2}$ has two boundaries; in this
case, we have
\begin{equation}
\label{Phi2}
\Phi_{2}=\int\int_{D_{2}}{\bf B}_{E}\cdot d{\bf S}_{2}=\oint_{\Gamma_{L}}{\bf A}_{E}
\cdot d{\bf\ell}-\oint_{\Gamma_{R}}{\bf A}_{E}\cdot d{\bf\ell}=0\ ,
\end{equation}
where we used the property (\ref{propr}).

The total magnetic flux through the whole region $D$ is
\begin{equation}
\label{Phi}
\Phi=\Phi_{1}+\Phi_{2}=\pi BR^{2}\ .
\end{equation}

With the appropriate choice of the $\gamma$-parameter, as made by
Aharonov and Bohm \cite{AB}, in equation (\ref{defA}) we set
\begin{equation}
\label{escolhagamma}
\gamma={BR^{2}\over 2}\ ;
\end{equation}
we have from (\ref{propr}),
\begin{equation}
\oint_{\Gamma_{I}}{\bf A}_{I}\cdot d{\bf\ell}=\oint_{\Gamma_{L}}{\bf A}_{E}\cdot
d{\bf\ell}=\oint_{\Gamma_{R}}{\bf A}_{E}\cdot d{\bf\ell}=\pi BR^{2}\ .
\end{equation}
With this peculiarity, and equations (\ref{Phi1}), (\ref{Phi2}) and (\ref{Phi}),
we can write
\begin{equation}
\label{fluxoGreen}
\Phi=\int\int_{D}{\bf A}\cdot d{\bf S}=\oint_{\Gamma_{L}}{\bf A}_{E}\cdot d{\bf\ell}=
\oint_{\Gamma_{L}}{\bf A}\cdot d{\bf\ell}=\pi BR^{2}\ .
\end{equation}

Equation (\ref{fluxoGreen}) has an interesting interpretation: it
is like we could apply  Stokes' Theorem to the potential
\begin{equation}
\label{Aescolhido}
{\bf A}(\rho,z,\phi)=\cases{{B\rho\over 2}{\hat\phi}, \rho<R\cr\cr
{BR^{2}\over 2\rho}{\hat\phi}, \rho>R}\ ,
\end{equation}
discarding the fact that it is not a $C^{1}$ field in the region $D$; we simply considered
the integral along the exterior path $\Gamma_{L}$.

We would like to emphasize that the value of the
$\gamma$-parameter (\ref{escolhagamma}) is only a choice, it is
not an imposition of Stokes' Theorem. Another important point to
notice is that, with this choice of $\gamma$, we can define a
continuous potential (\ref{Aescolhido}) at  $\rho=R$, (if we
consider it defined in this region) but its first derivative is
still discontinuous, and consequently, it is not a
$C^{1}$-field\footnote{In the spirit of the theory of
distributions, a vector potential of the form (\ref{defA}) can
produce a magnetic delta field on the solenoid ($\rho=R$). In a
classical context, it is not a problem, since the magnetic field
is given by (\ref{defB}) in the physical regions \cite{nosso}. A
similar fact happens in the Dirac quantization, where, on the
Dirac's string, the magnetic field diverges \cite{Olive,nosso}.}.

In the context of Quantum Mechanics, the situation is not so
simple. As pointed out by Aharonov and Bohm \cite{AB}, in spite of
the magnetic field ${\bf B}_{E}$ in the exterior region to be
zero, a vector potential of the form (\ref{defA}) can yield
measurable physical phenomena, even with its rotational being
zero. These phenomena, usually known as Aharonov-Bohm effects,
have their explanations based on the fact that, by virtue of
equation (\ref{defA}), when we consider wave functions in the
exterior region of the solenoid, these ones become no more
single-valued functions, acquiring a phase factor when we take a
closed path around the solenoid; so, for an wave function $\psi$
outside the solenoid, we have
\begin{eqnarray}
\label{funcaodeonda}
\psi(r,\phi+2\pi,z)&=&\exp\Bigl(-iq\oint{\bf A}_{E}\cdot d{\bf\ell}\Bigr)\psi(r,\phi,z)
\cr\cr
&=&\exp{(-2\pi iq\gamma)}\psi(r,\phi,z)\ ,
\end{eqnarray}
where $q$ is the charge of the considered wave
function\footnote{These phenomena are also based on the fact that
the domain of the wave function (the exterior region) is not
simply-connected \cite{Rivers}.}.

The physical quantities introduced by the property
(\ref{funcaodeonda}) are proportional to the phase difference
$q\oint{\bf A}_{E}\cdot d{\bf\ell}=2\pi iq\gamma$ and are periodic
functions of the parameter $\gamma$, with period given by $1/q$.
If we have $q\oint{\bf A}_{E}\cdot d{\bf\ell}=2\pi iq\gamma=2\pi
N$, for $N=0,\pm1,\pm2,...$, the wave function
(\ref{funcaodeonda}) becomes single valued and we have no physical
effect.

In order to keep the vector potential continuous over the whole
space, and to keep the familiar expression (\ref{fluxoGreen}) for
Stokes' Theorem, as already said, Aharonov and Bohm took the
vector potential as being the one expressed in equation
(\ref{Aescolhido}), which turns equation (\ref{funcaodeonda}) into
the form
\begin{equation}
\label{diffase1}
\psi(r,\phi+2\pi,z)=\exp(-iq\pi BR^{2})\psi(r,\phi,z)\ ,
\end{equation}
where we used equation (\ref{fluxoGreen}).

Ever since, many other problems dealing with an infinitely long
solenoid, as discussed above, were studied in the literature, and
in all of them, it is considered the vector potential with the
form (\ref{Aescolhido}). As a consequence, the physical result
predictions for all of these problems are proportional to the
phase difference presented in (\ref{diffase1}).

Different choices for the value for the $\gamma$-parameter in
equation (\ref{defA}) would  give different physical results for
effects of the Aharonov-Bohm kind. In spite of this, Stokes'
Theorem and Quantum Mechanics do not state that the value of
$\gamma$ must be the one expressed in equation
(\ref{escolhagamma}); it was simply a choice.

Nowadays, there are experimental evidences for the existence of
Aharonov-Bohm effects, in agreement with the value
(\ref{escolhagamma}). What we wish to explore in this text is the
fact that (\ref{escolhagamma}) is not the only choice that can be
done for the $\gamma$-parameter according to the experiments.
There is still a certain freedom in the value of $\gamma$. In
order to study this point and its consequences, instead of
(\ref{escolhagamma}), let us write:
\begin{equation}
\label{gammacomkappa}
\gamma={BR^{2}\over 2}+\kappa\ ,
\end{equation}
where $\kappa$ is any real constant. Inserting (\ref{gammacomkappa})
into equation (\ref{defA}), and using (\ref{funcaodeonda}), we have
\begin{equation}
\label{diffase2}
\psi(r,\phi+2\pi,z)=\exp(-iq\pi BR^{2})\exp(-2\pi iq\kappa)\psi(r,\phi,z)\ .
\end{equation}
We do not expect that the choice (\ref{gammacomkappa}) gives physical results different
from the ones presented in the literature, and obtained, with the choice
(\ref{escolhagamma}) those for which there are experimental verifications. To
assure this, the phase difference of the wave function in (\ref{diffase2}) must be
the same phase difference of the equation (\ref{diffase1}). This condition is satisfied
by taking
\begin{equation}
\label{condicaokappa}
q\kappa=n_{q,\kappa}\ ,
\end{equation}
where $n_{q,\kappa}$ is an integer number that depends on the charge $q$ of
the wave function, and on the parameter $\kappa$. In order to get the possible values of
$\kappa$, we consider the expression above for the case of the electron charge, $e$.
The result is
\begin{equation}
\label{valorkappa}
\kappa={n_{e,\kappa}\over e}={N_{\kappa}\over e}\ ,
\end{equation}
where $N_{\kappa}$ is an integer that depends only on $\kappa$.

Using equations (\ref{condicaokappa}) and (\ref{valorkappa}), we
have that any given value of electric charge must satisfy the
condition
\begin{equation}
q={n_{q,\kappa}\over N_{\kappa}}e\ .
\end{equation}
The only way to avoid a $\kappa$-dependence for the charge $q$ is
to take $n_{q,\kappa}=n_{q}n_{\kappa}$ and
$N_{\kappa}=Nn_{\kappa}$, where $n_{q}$ and $n_{\kappa}$ are
integers that depend, respectively, on $q$ and $\kappa$, and $N$
is an integer constant. This finally leads to
\begin{equation}
\label{resultado}
q={n_{q}\over N}e\ ,
\end{equation}
that is, any value of electric charge $q$ is an integer multiple of a
given fraction of the electron charge $e$.

It is interesting to notice that the result (\ref{resultado}) is
consistent with the existence of anti-particles and with the
quarks charges. From equation (\ref{valorkappa}), we can see that
the possible values of the $\kappa$-parameter, that was completely
free at the classical level, become quantized in the quantum
context.

The result (\ref{resultado}) could be proposed by means of a
slightly different point of view. Equation (\ref{gammacomkappa})
is equivalent to a gauge transformation of the form
\begin{equation}
\label{transformacao}
{\bf A}\rightarrow{\bf A}+{\kappa\over\rho}{\hat\phi}\ ,
\end{equation}
for the vector potential (\ref{Aescolhido}) in the exterior region
of the solenoid. This gauge transformation is not in disagreement
with Stokes' Theorem and with Quantum Mechanics. In the context of
the Classical Electrodynamics, it will produce no effect, but in
the context of Quantum Mechanics, it will produce a phase
difference of the form (\ref{diffase2}) in a wave function outside
the solenoid. As a consequence, we will have, in the problems of
the Aharonov-Bohm type, a correction to the physically measurable
quantities due only to the gauge transformation
(\ref{transformacao}). The only way to avoid physical dependences
on the gauge parameter $\kappa$ is to have the condition
(\ref{condicaokappa}). With the previous argument, we are taken to
equation (\ref{resultado}).

A similar gauge transformation to (\ref{transformacao}) occurs in
the Dirac's quantization and it is produced in considering that
the vector potentials generated by two Dirac's strings, related to
the same magnetic monopole, but located in different regions of
space, can be related by a gauge transformation of the form
(\ref{transformacao}).

In our case, the transformation (\ref{transformacao}) occurs
naturally, due only to the fact that it is well-defined in the
exterior region of the solenoid because the $z$-axis is excluded
from this region (what makes it a region with a non trivial
geometry). A similar analysis could be carried out from other
situations where we have non-trivial geometries.

As a last comment, we would like to say that the freedom in the
$\gamma$-parameter in equation (\ref{defA}), or equivalently, the
gauge transformation (\ref{transformacao}), was used, some years
ago, to argue that the Aharonov-Bohm does not exist \cite{1}. This
idea was refused by some authors with the argument that the
transformation (\ref{transformacao}) was in disagreement with the
Stokes' Theorem \cite{2}.

\section{Conclusions and final remarks}

We have shown that Stokes' Theorem allows us to have, in the
Aharonov-Bohm experiment, a vector potential outside the solenoid,
whose circulation around a closed path is not equal to the flux of
the magnetic field through a surface whose boundary is the
considered path. We have shown that, when we consider this
possibility in the context of Quantum Mechanics, by choosing
conveniently the vector potential in the exterior of the solenoid,
we are taken straight to a quantization condition for the values
of the electric charges, without appealing to the existence of
magnetic monopoles.

The convenient choice for the vector potential quoted above can be
interpreted as a gauge transformation, which has the form of an
Aharonov-Bohm potential. The quantization condition attained
thereby is consistent with the quark electric charges and with the
existence of antiparticles.

We think that these results can be extended to the case of
non-Abelian fields, where a quantization condition for the color
charges must be obtained too.

\section*{Acknowledgements}

The authors would like to thank M. V. Cougo-Pinto, C. Farina, H.
Boschi-Filho and F.E. Barone for useful discussions. Professor R.
Jackiw is kindly acknowledged for relevant comments, and Professor
Nathan Lepora for the help with relevant references. F.A.B. thanks
FAPERJ for the invaluable financial help.



\end{document}